\documentclass[a4paper,11pt]{article}
\usepackage{jinstpub} 
\usepackage{lineno}
\nolinenumbers

\usepackage{svg}

\title{\boldmath Normalizing Flows for Domain Adaptation when Identifying $\Lambda$ Hyperon Events}

\author[a,1]{R. Kelleher\note{Corresponding author.}}
\author[a,b]{A. Vossen}
\affiliation[a]{Department of Physics, Duke University, \\ 120 Science Drive, Durham, NC 27708, USA}
\affiliation[b]{Thomas Jefferson National Accelerator Facility, \\ 12000 Jefferson Ave., Newport News, VA 23606, USA}

\emailAdd{rowan.kelleher@duke.edu}

\abstract{This study focuses on the application of a normalizing flow as a method of domain adaptation when classifying physics data. Normalizing flows offer a way to transform data points between two different distributions. The present study investigates a novel method of transforming latent representations of physics data to a normal distribution and then to a physics distribution again. The final distribution models a simulated distribution. After being transformed, the data can be classified by a neural network trained on labeled simulation data. The present study succeeds in training two normalizing flows that can transform between data (or simulation) and a Gaussian distribution.}


\begin{document}
\maketitle
\flushbottom

\section{Introduction}
\label{sec:intro}
In semi-inclusive deep inelastic scattering (SIDIS) experiments~\cite{Spinstructure}, electrons are scattered off of protons to probe the spin structure of the proton. $\Lambda$ hyperons are produced in the scattering experiment at the CLAS12 detector~\cite{CLAS12}, but have a lifetime too short to detect directly in the detector. $\Lambda$ hyperons primarily decay into a proton and negatively charged pion; events where a proton and $\pi^-$  are measured in the final state may have produced a $\Lambda$. However, protons and $\pi^-$ may be produced by other reactions, meaning events where a $\Lambda$ is produced must be distinguished from those where there are only background processes. A signal fit can be applied with a peak at the nominal $\Lambda$ mass of 1.1157 GeV. Although this process can work well enough, there is a prominent background for these events at CLAS12, meaning that the signal fit alone produces a relatively low signal to background ratio. Improving the signal to background ratio could help increase confidence in the signal and background separation of the fit, reducing statistical uncertainty. 

One approach to improving signal extraction consists in training a classifier on kinematic variables related to $\Lambda$ event candidates so that the classifier can identify which events contained a $\Lambda$ decay. The CLAS12 detector cannot record data for particles that decay before interacting with the detector volumes. Instead, particles that decay quickly produce decay products that are measured. These products can help us understand the processes that occurred during a collision, but do not provide enough information to definitively identify which decay products resulted from which parent particles. Thus, measured data cannot be labeled as containing $\Lambda$ hyperons, meaning they cannot be used for training. Simulated data can contain all information about events, allowing it to be labeled. Monte-Carlo event generation (MC)~\cite{MC_Pepsi_lund} was used to produce SIDIS events that simulate the physics that occurs in the CLAS12 detector, which is used in the present study. Training a classifier on this simulation data provides a tool to identify which simulated events contain a $\Lambda$, but this does not directly translate to datasets composed of measured events. Differences between the simulated and measured data causes the classifier to struggle to accurately classify events from the measured dataset. Previous work has been done in Ref.~\cite{McEneaney_2023} where a domain adversarial graph neural network was employed to improve the classification process and the signal extraction. The present study builds on the graph neural network (GNN) implementation by adding a domain adaptation stage in the classification process between the GNN and the classification network. The domain adaptation is implemented via the use of two normalizing flow networks that transform the dataset between data and simulation distributions.

\section{Normalizing Flows}
\label{sec:normalizing_flows}
Normalizing flows are a class of neural networks that transform data points between a simple base distribution and a complex distribution. This process is achieved through the layering of many invertible, differentiable mappings. When used for generation tasks, samples are produced by the output of these mappings. Both the density of the sample under the base distribution and the change of volume originating from the transformation can be calculated. The product of this density and volume can be treated as a likelihood, and the negative likelihood can be thought of as loss~\cite{Normalizing_flow_intro}.

The change of variables formula is central to the normalizing flow architecture as it enables the computation of the probability density function (PDF) of a complex distribution. We can start with a base distribution Z for which we know the PDF (in D dimensions where our input is of dimension D), then find a function that transforms Z to a distribution X with an unknown but desired PDF, $p_X(x)$. With both the base distribution and transformation, we can compute the PDF of Z, $p_Z(z)$ and sample from it. The transformation must be bijective, however, to allow for computations of its inverse. Many individual bijections are chained together to create a function powerful enough to transform X to a Gaussian.

The functions that describe the transformation from Z to X can be learned through log-likelihood maximization (or negative log-likelihood minimization). This likelihood is computed by taking the logarithm of the change of variables formula: 
\begin{equation}\log{p_X(x)}=\log{p_Z(f(x))} + \log{ \Bigl| det \Bigl( \frac{\partial f(x)}{\partial x^T}\Bigr) \Bigr|} \end{equation}\label{eqn:changeofvariables}

The parameters of Z and $f(x)$ are trained according to this equation.

\subsection{Architecture}
\label{sec:architecture}
The present study utilizes a flow model based on the RealNVP architecture introduced by Ref.~\cite{dinh2017density}. The RealNVP architecture implements the transformation as a composition of coupled functions, where each function scales (with scaling function $s(x)$) and translates (with translation function $t(x)$) the input. To improve computation efficiency, the architecture uses coupling functions where the input vector is split into two segments. The output of the first segment is set as equal to the input of the first segment; the output of the second segment is parameterized by the first segment:
\begin{equation}
\label{eqn:coupling}
\begin{aligned}
y_{1:d} &= x_{1:d}\\
y_{d+1:D} &=x_{d+1:D}\cdot exp(s(x_{1:d})) + t(x_{1:d})
\end{aligned}
\end{equation}

This coupling method improves the computation efficiency for the log-determinant as the coupling forces the transformation's jacobian to be triangular. The determinant of a triangular matrix is the product of its diagonal entries; because the diagonal entries consist only in the scale portion of the transformation, and the scaling is exponential, the exponentials can be summed. Now, the determinant computation consists only of a summation, which is very favorable for the computation time of the training process.

\section{Model Implementation}
\label{sec:model_implementation}
The normalizing flow model used for the present study implements the normalizing-flows python package~\cite{Stimper2023}. The normalizing-flows package provides implementations of the affine coupling layers described in section~\ref{sec:architecture}. A model was constructed with 32 masked-affine layers; bit-masking allows for the coupling between different segments of the input as described in section~\ref{sec:architecture}, condensing eqn. \ref{eqn:coupling} into a single parametric equation with a bit-mask b (given in Ref.~\cite{dinh2017density}):
\begin{equation}
\label{eqn:parametric}
y = b \cdot x + (1 - b) \cdot (x \cdot exp(s(b \cdot)) + t(b \cdot x) )
\end{equation}
The bit-mask is alternated with each layer to ensure different input dimensions are transformed every other layer.

The scale ($s$) and translation ($t$) functions used in eqn. \ref{eqn:parametric} are implemented as multi-layer perceptrons (MLPs) using the pytorch python package~\cite{PyTorch} in the normalizing-flows package. These MLPs utilize input and output dimensions of 71 to match the output dimension of the GNN. Each MLP has two hidden layers with dimensions of 142. The normalizing-flows package provides functions for calculating negative log-likelihood as explained in section~\ref{sec:architecture}, as well as an implementation for back propagating the loss. These functions are used for training the model. 

Each normalizing flow model was trained on batches of 100 events. For each batch, the training loop: transforms the batch of events according to the current parameters; calculates the likelihood that the transformed events were drawn from the base distribution; updates the model's parameters through gradient descent. A validation dataset was used to determine when to stop training. A testing dataset was used at the end of training to ensure the models did not overfit or diverge.
\section{Application}
\label{sec:application}
\begin{figure}[h]
    \centering
    \includegraphics[width=0.7\textwidth]{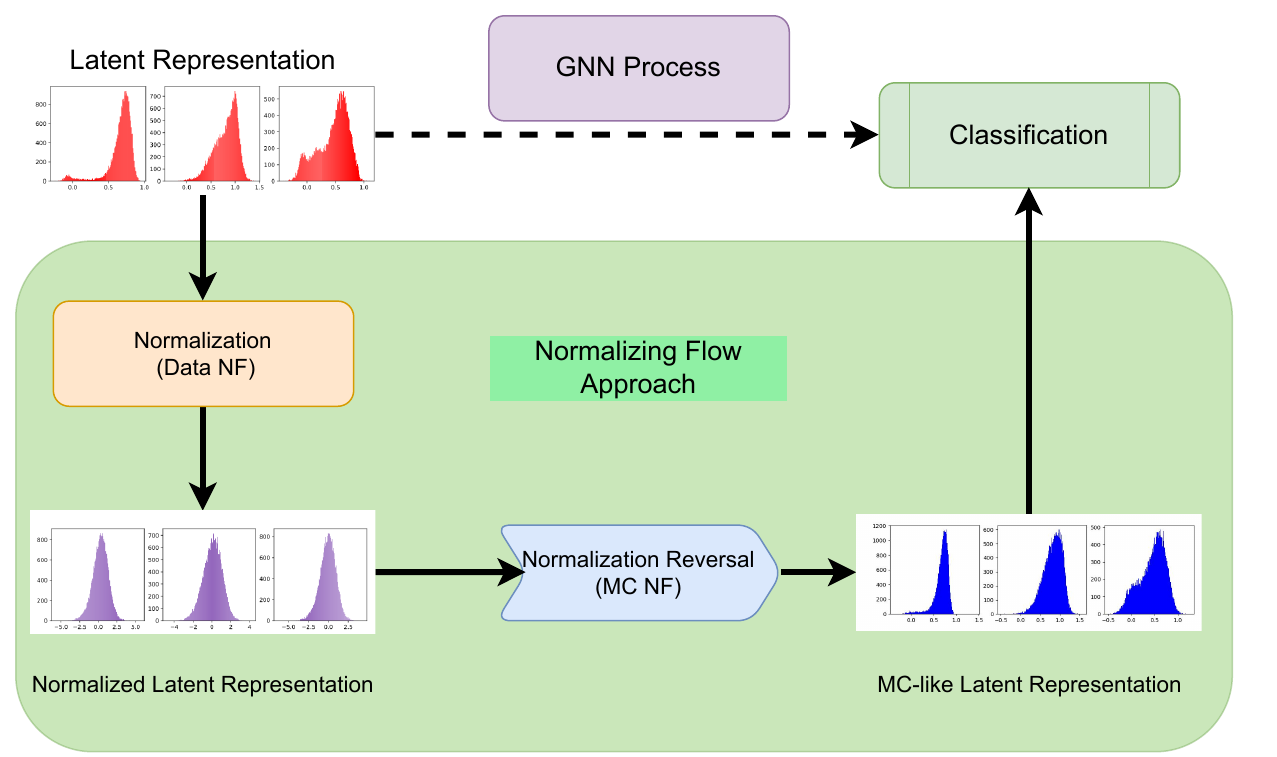}
    \caption{The transformation process inserted between latent space extraction and classification steps.}
    \label{fig:NF_workflow}
\end{figure}

Our application of normalizing flows as a form of domain adaptation aims to improve the performance of a classifier trained to identify simulated $\Lambda$ events. Previous studies have investigated the use of normalizing flows for domain adaptation in high energy physics contexts, such as Ref.~\cite{algren2023flow}. We introduce a novel approach where two models are placed back-to-back to transform between two unknown distributions. Data is first transformed to a normal distribution via a forward pass with a data-network. Then, the normalized data would be passed backwards through a simulation-network such that it ends up "looking like" the simulation data that the classifier was trained on. By using an intermediate normal distribution, we can transform between the data and simulation distributions. Figure~\ref{fig:NF_workflow} visualizes the process.

For this strategy two different networks were trained separately: a data-network was trained to normalize measured data; a simulation-network was trained to normalize simulation data. Because the networks represent bijective functions, they can be reversed to turn normalized samples into simulation-like samples. The data-network was trained over 6 epochs while the simulation-network was trained over 11 epochs (due to the size difference in the datasets). 

An additional study using the normalizing flow model described in section~\ref{sec:model_implementation} investigated the model's ability to reverse distortion in physics data distributions. In this study, the data was transformed without feature extraction (via a GNN for instance), and the results are discussed in a separate article~\cite{SPINproceedings}.
\section{Results}
\label{sec:results}
\begin{figure}[b]
    \centering
    \includegraphics[width=0.47\textwidth]{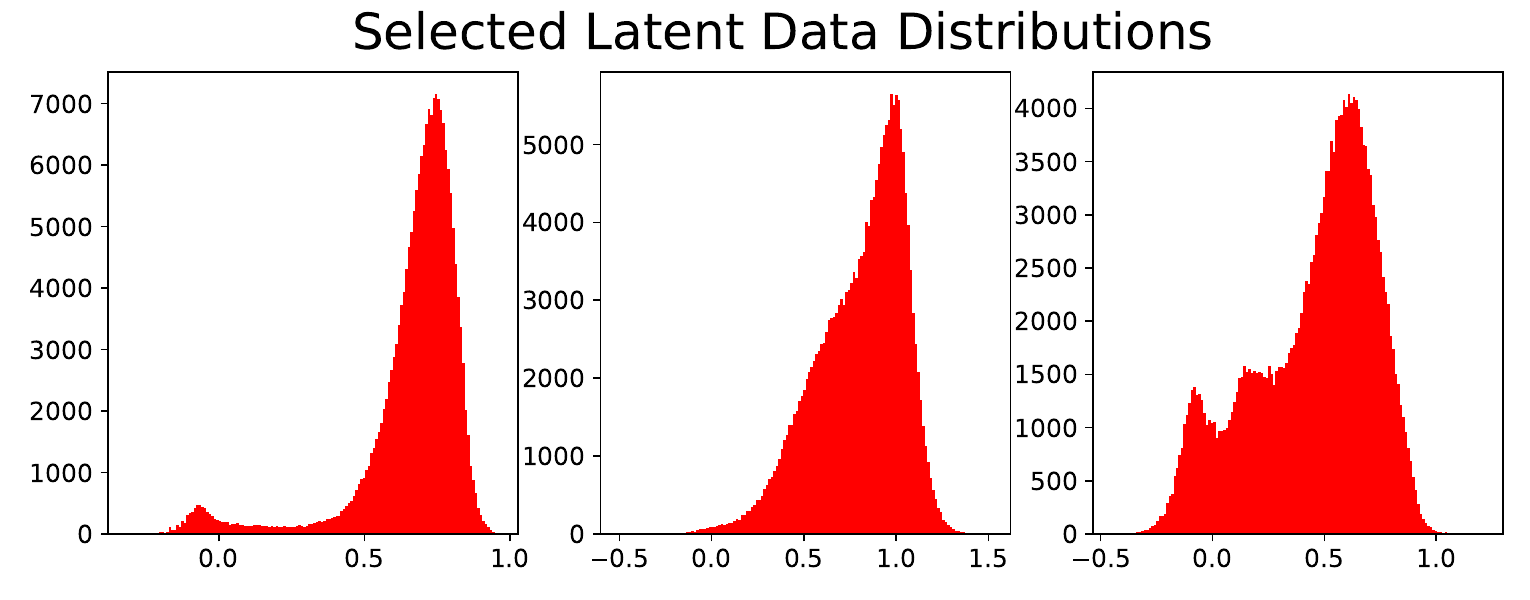}
    \qquad
    \includegraphics[width=0.47\textwidth]{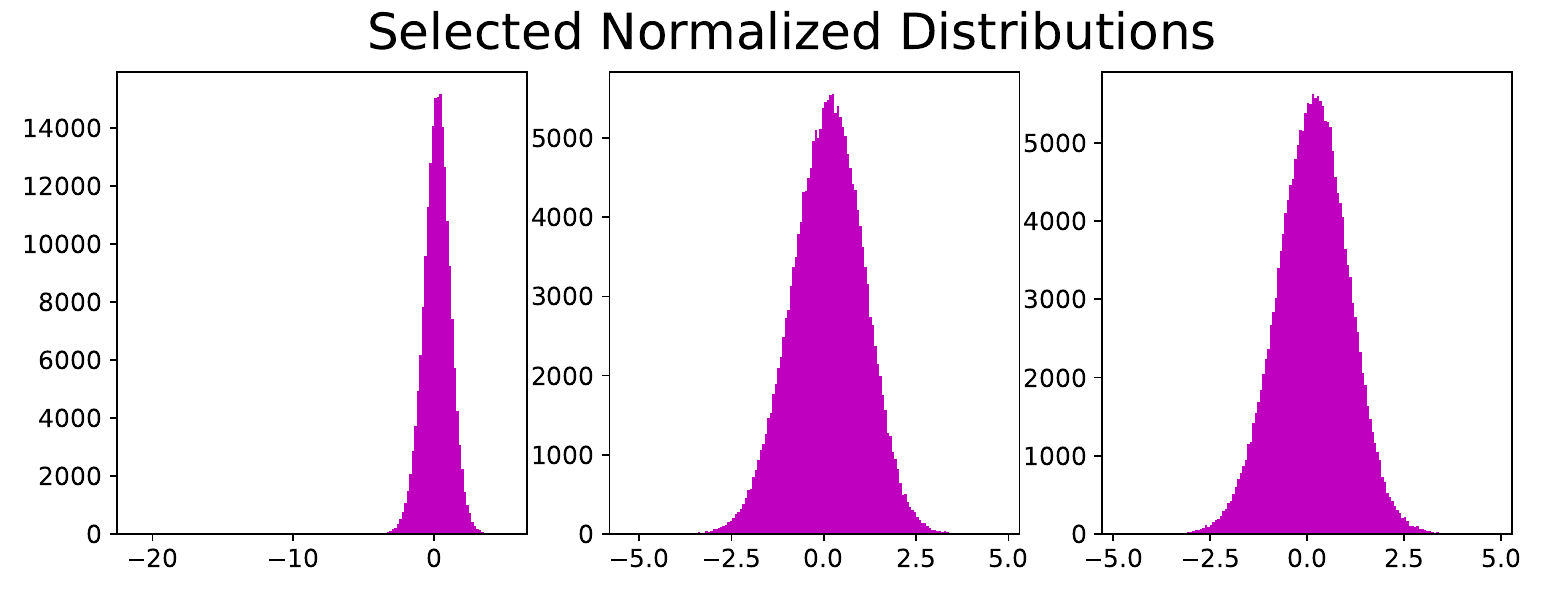}
    \caption{Histograms are shown for several dimensions of the latent representation of data before (left) and after normalization (right).}
    \label{fig:transform_hist}
\end{figure}

\begin{figure}[h]
    \centering
    \includegraphics[width=0.45\textwidth]{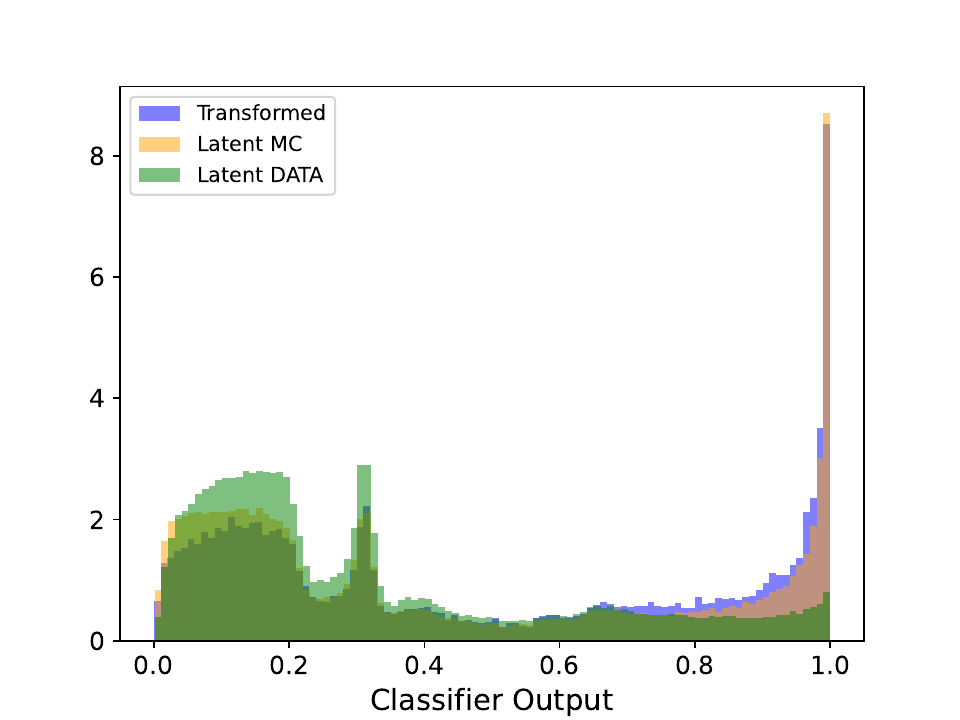}
    \qquad
    \includegraphics[width=0.45\textwidth]{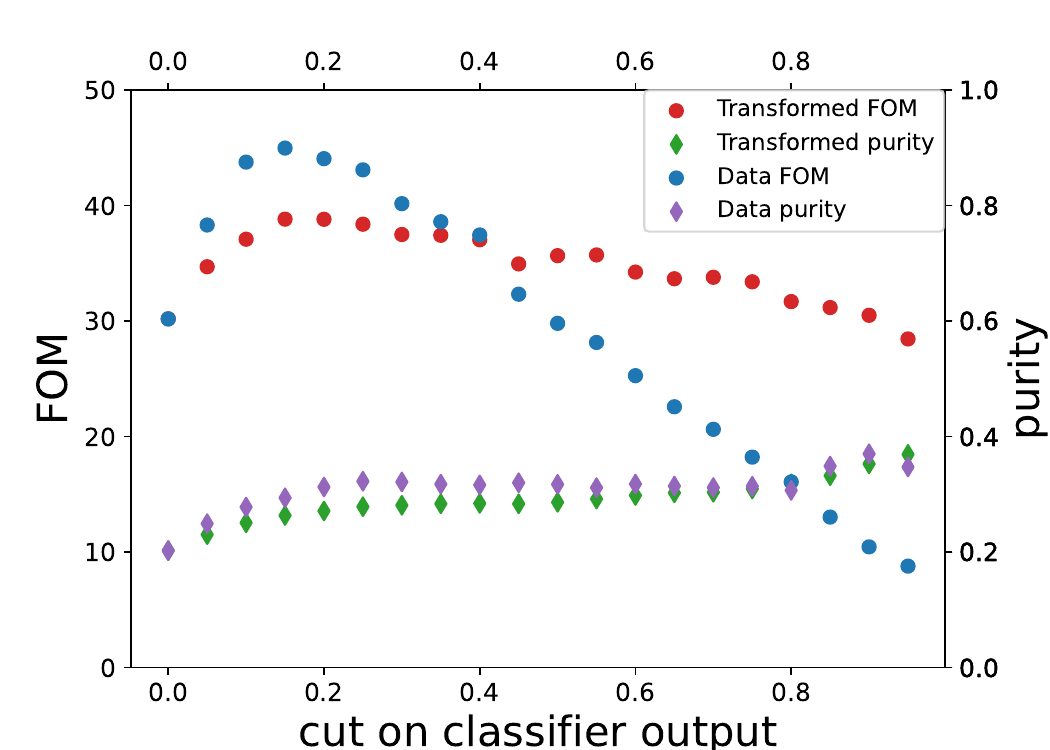}
    \caption{The classifier output for the data, simulation, and transformed data are shown as well (left). The figure of merit and purity are plotted for both transformed and latent data fits (right).}
    \label{fig:Classifier_and_FOM}
\end{figure}

The models were successful in learning to normalize their respective inputs, as shown in Fig.~\ref{fig:transform_hist}. The validation loss appeared to match the training loss well, suggesting we did not encounter over-fitting. The normalized distributions appear to follow a normal distribution, however we can notice that some dimensions have distributions that differ from the expected result. Some distributions are not centered around 0 and some are skewed.
The classifier output for the simulation, the data, and the transformed data are shown in Fig.~\ref{fig:Classifier_and_FOM}. The classifier output for the transformed data matches that of the simulation much better than that of the data, suggesting that the domain adaptation helped align the data with the inputs that the classifier was trained on. Furthermore, the figure of merit ($N_{signal} / \sqrt{N_{tot}}$) and purity ($N_{sig} / N_{tot}$) for the data and transformed data are shown in Fig.~\ref{fig:Classifier_and_FOM}, where $N_{tot}$ represents the total number of events being fit and $N_{sig}$ represents the number of events that reside inside the fit.

The figure of merit appears flatter when transformed with the flow model. The flatness of the curve may be desirable as it allows for cuts to be made almost anywhere on the curve without sacrificing the figure of merit.

\section{Conclusions}
In this study, we have implemented a normalizing flow neural network to improve the ability of a classifier network to classify measured physics data. Flow based networks can be used to learn unknown probability distributions of measured data via the training of a transformation between the unknown distribution and a known distribution. The present study has succeeded in producing a normalizing flow data model that can learn the transformation between a latent representation of physics data and a normal distribution as shown in Fig.~\ref{fig:transform_hist}.

Our study utilizes a novel approach where two normalizing flow models are used back-to-back to transform between two unknown distributions. We achieve this by normalizing the latent data via a model trained on measured data, then transforming in the reverse direction via a model trained to normalize simulated data. We have shown that performing this transformation can improve the classification of of physics events, creating greater generalizability when cutting on the classifier output. Future work could be done to investigate different types of flow models other than the realNVP model implemented in this study.
\clearpage
\bibliographystyle{JHEP}
\bibliography{biblio.bib}
\end{document}